\documentclass[a4paper,11pt]{article}
\usepackage{moriond,epsfig}
\usepackage{units}

\def\be{\begin{equation}}
\def\ee{\end{equation}}
\def\bea{\begin{eqnarray}}
\def\eea{\end{eqnarray}}

\newcommand{\bq}{\begin{equation}}
\newcommand{\eq}{\end{equation}}

\newcommand{\bqq}{\begin{eqnarray}}
\newcommand{\eqq}{\end{eqnarray}}

\newcommand{\cmsNN}{\sqrt{s_{NN}}}

\newcommand{\expval}[1]{\langle #1 \rangle}

\newcommand{\etain}[1]{$|\eta|$~$<$~$#1$}

\newcommand{\pt}{\ensuremath{p_T} }
\newcommand{\pta}{\ensuremath{p_{T, \rm assoc}}}
\newcommand{\ptt}{\ensuremath{p_{T, \rm trig}}}

\newcommand{\Dphi}{\Delta\phi}
\newcommand{\Deta}{\Delta\eta}
\newcommand{\Ntrig}{N_{\rm trig}}
\newcommand{\Nassoc}{N_{\rm assoc}}
\newcommand{\icp}{I_{CP}}
\newcommand{\iaa}{I_{AA}}
\newcommand{\iaapythia}{I_{AA,Pythia}}

\newcommand{\figref}[1]{Fig.~\ref{#1}}

\newcommand{\bfigFullPage}{\begin{figure} \begin{center} \vspace{0pt}}
\newcommand{\bfig}[1][t!]{\begin{figure}[#1] \begin{center}}
\newcommand{\efig}{\end{center} \end{figure}}
\newcommand{\btab}[1][t!]{\begin{table}[#1] \begin{center}}
\newcommand{\etab}{\end{center} \end{table}}

\begin{document}
\vspace*{4cm}
\title{Measurement of jet quenching with $\icp$ and $\iaapythia$ in Pb-Pb collisions at $\cmsNN = \unit[2.76]{TeV}$ with ALICE}

\author{Jan Fiete Grosse-Oetringhaus for the ALICE collaboration}

\address{CERN, 1211 Geneva 23}

\maketitle\abstracts{
This paper discusses the measurement of $\icp$ and $\iaapythia$ with ALICE (A Large Ion Collider Experiment). An away-side suppression is found expected from in-medium energy loss. Further, and unexpected, a near-side enhancement is seen which has not been reported by previous experiments at lower energies.}


\vspace{-0.4cm}
The objective of the study of ultra-relativistic heavy ion-collisions  is the characterization of the quark--gluon plasma, the deconfined state of quarks and gluons. 
Recent measurements by ALICE indicate that in central Pb-Pb collisions at the LHC unprecedented color charge densities are reached. For example, the suppression of charged hadrons in central Pb-Pb collisions at $\cmsNN = \unit[2.76]{TeV}$ expressed as the nuclear modification factor $R_{\rm AA}$ as a function of transverse momentum ($\pt$) reaches a value as low as 0.14.\cite{raa}

Di-hadron correlations allow for the further study of in-medium energy because for most pairs of partons scattered in opposite directions, one will have a longer path through the medium than the other. Thus, two-particle correlations can be used to study medium effects without the need of jet reconstruction.
In such studies the near-side (particles found close to each other in azimuthal angle) and the away-side (particles found at azimuthal angles different by about $\pi$) yields are compared between central and peripheral events ($\icp$) or studied with respect to a pp reference ($\iaa$). Previous measurements at RHIC have shown a significant suppression of the away-side yield consistent with a strongly interacting medium.\cite{iaaphenix,iaastar} On the near-side no significant modifications have been observed at high $\pt$. Such analysis usually require the subtraction of non-jet correlations, e.g. flow, which are present in A+A collisions but not in pp collisions and therefore have influence on the extracted yields. This analysis chooses a $\pt$-region where the jet peak is the dominant correlated signal and thus the influence of non-jet correlations is small.

\vspace{-0.2cm}
\section{Detector and Data Sample}

The ALICE detector is described in detail elsewhere.\cite{alice} For the present analysis the Inner Tracking System (ITS) and the Time Projection Chamber (TPC) are used for vertex finding and tracking. The TPC has a uniform acceptance in azimuthal angle and a pseudorapidity coverage of \etain{0.9}. The uniform acceptance results in only small required acceptance corrections. Forward scintillators (V0) are used to determine the centrality of the collisions.

About 12 million minimum-bias events recorded in fall 2010 have been used in the analysis. Good-quality tracks are selected by requiring at least 70 (out of 159) associated clusters in the TPC, and a $\chi^2$ per space point of the momentum fit smaller than 4. In addition, tracks are required to originate from within $\unit[2-3]{cm}$ of the primary vertex.

\section{Analysis}

The quantity which is obtained in this analysis is the associated per-trigger yield as function of the azimuthal angle difference:
\bq
  \frac{dN}{d\Dphi}(\Dphi) = \frac{1}{\Ntrig} \frac{d\Nassoc}{d\Dphi} \label{phi_yield}
\eq
where $\Ntrig$ is the number of trigger particles to which $\Nassoc$ particles are associated at $\Dphi = \phi_{\rm trig} - \phi_{\rm assoc}$.
We measure this quantity for all pairs of particles where $\pta < \ptt$ within \etain{0.8} and normalize by $\Deta = 1.6$. Due to the flat acceptance in $\phi$ no mixed-event correction is needed. The per-trigger yield is extracted in bins of $\ptt$ and $\pta$.

\vspace{-0.2cm}
\paragraph{Pedestal Subtraction}

\bfig
  \includegraphics[width=0.75\linewidth,trim=0 5 0 5,clip=true]{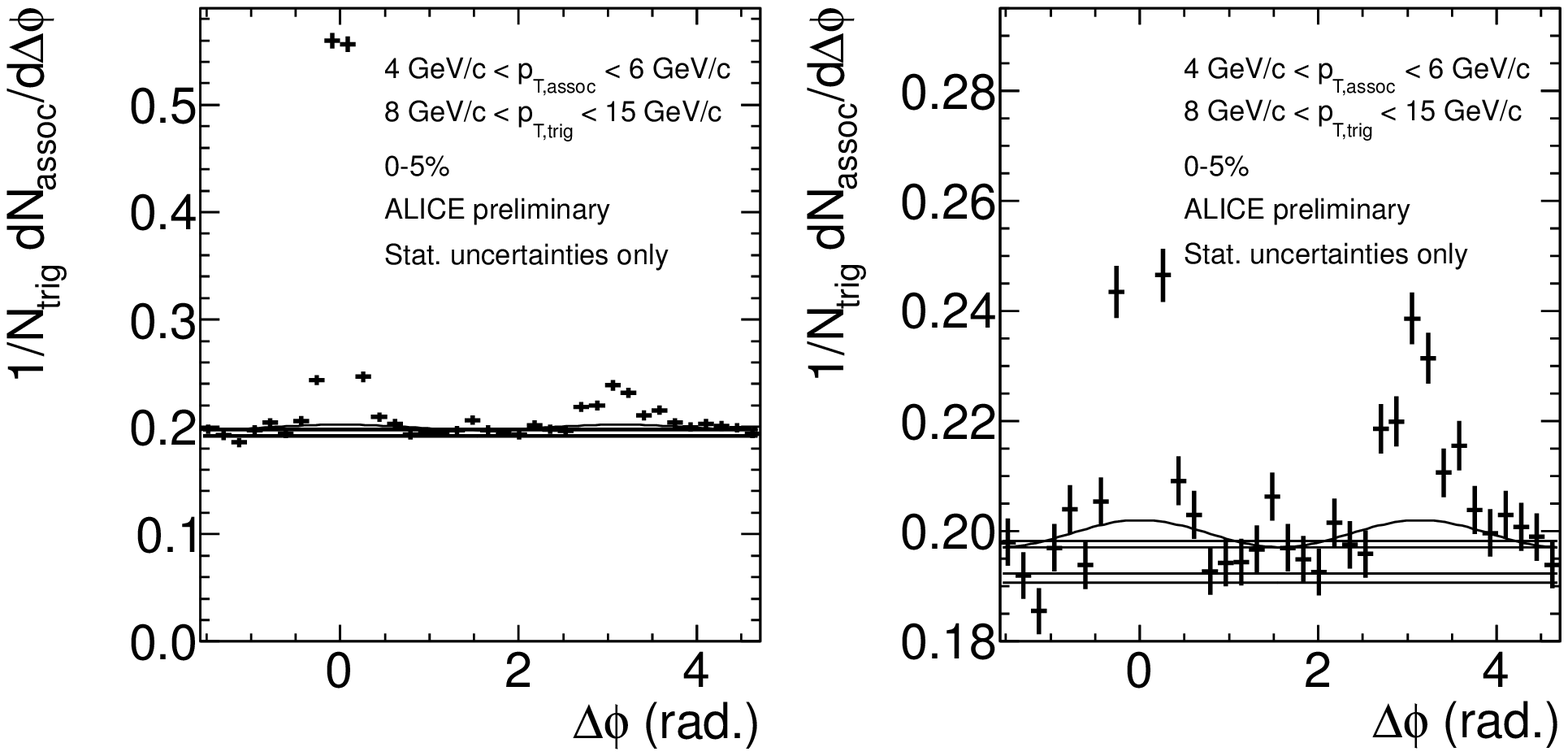}
  \caption{\label{fig_pedestal} Per-trigger yield in an example bin: the right panel shows a zoom of the left panel. Indicated are the determined pedestal values (horizontal lines) and the $v_2$ component ($\cos 2 \Dphi$ term). For details see text.}
\efig

To remove uncorrelated background from the associated yield, the pedestal value needs to be determined. This is done by fitting the region close to the minimum of the $\Dphi$ distribution ($\Dphi \approx \pm \frac{\pi}{2}$) with a constant and using this value as pedestal (zero yield at minimum -- ZYAM). One cannot exclude a correlated contribution in this region (e.g. from 3-jet events), and we do not claim that we only remove uncorrelated background. Instead we measure a yield with the prescription given here.
To estimate the uncertainty on the pedestal determination, we use four different approaches (different fit regions as well as averaging over a number of bins with the smallest content). \figref{fig_pedestal} shows the per-trigger yield for an example bin. The horizontal lines indicate the determined pedestal values; their spread gives an idea of the uncertainty. Also indicated is a background shape considering $v_2$. The $v_2$ values are taken from an independent measurement (a measurement of $v_2$ at high $\pt$ similar to \cite{flow}. For the centrality class $60-90\%$ no $v_2$ measurement was available, therefore, as an upper limit, $v_2$ is taken from the $40-50\%$ centrality class as it is expected to reduce towards peripheral collisions). For a given bin the $v_2$ background is $2\expval{v_{\rm 2,trig}}\expval{v_{\rm 2,assoc}} \cos 2 \Dphi$ where the $\expval{...}$ is calculated taking into account the $\pt$ distribution of the trigger and associated particles. The yields are then calculated with and without removing the $v_2$ component.
Subsequently to the pedestal (and optionally $v_2$) subtraction, the near and away side yields are integrated within $\Dphi$ of $\pm 0.7$ and $\pi \pm 0.7$, respectively.

\vspace{-0.2cm}
\paragraph{Systematic Uncertainties}
The influence of the following effects has been studied and considered for the systematic uncertainty on the extracted yields: detector efficiency and two-track effects, uncertainties in the centrality determination, $\pt$ resolution, the size of the integration window for the near and away-side yield as well as uncertainties in the pedestal determination. The last mentioned item has the largest contribution (7-20\%) to the systematic uncertainties on $\icp$ and $\iaapythia$.

\vspace{-0.2cm}
\paragraph{Results}

\bfig
  \includegraphics[width=0.9\linewidth,trim=0 15 0 12,clip=true]{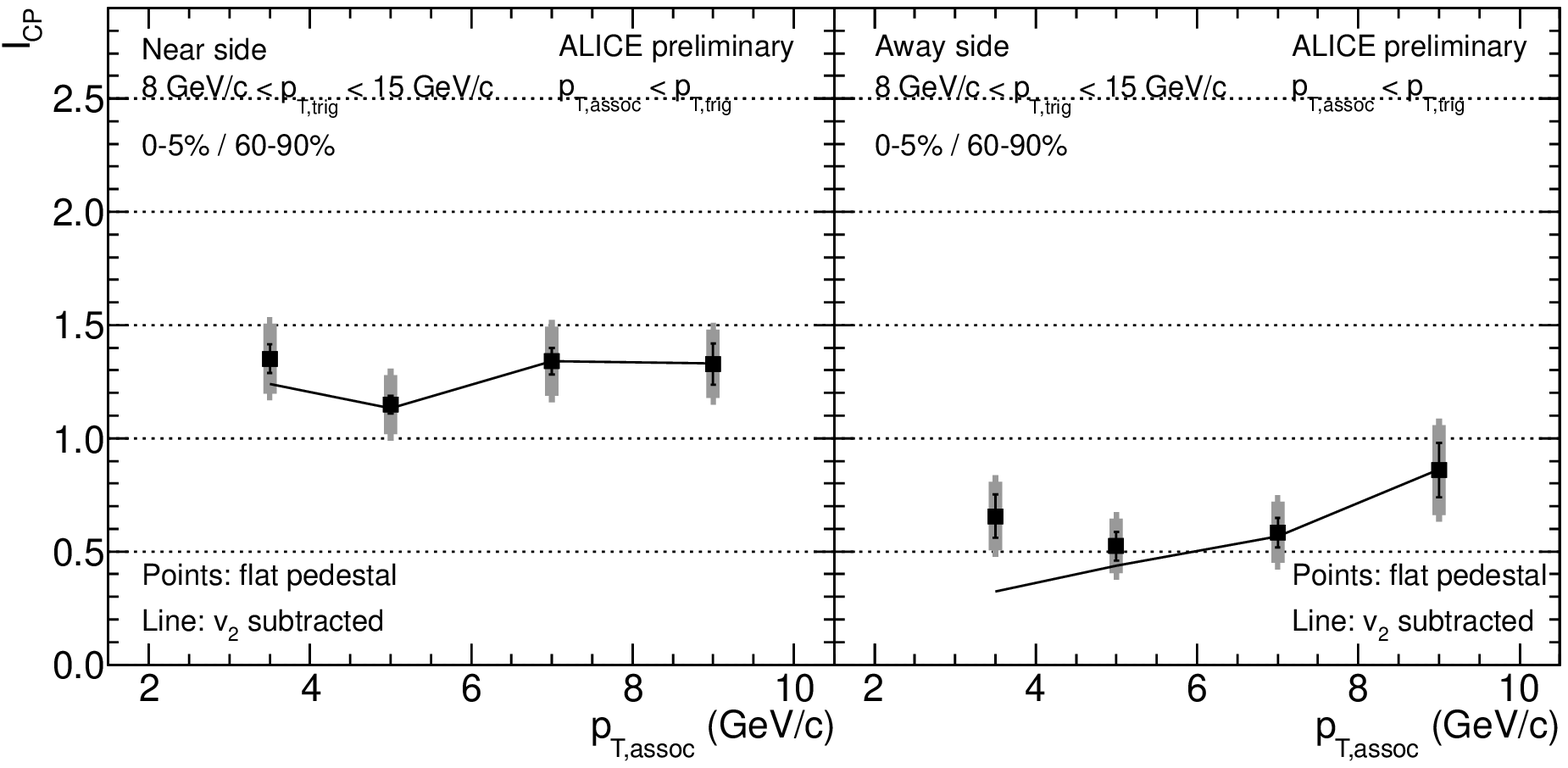}
  \caption{\label{fig_icp} $\icp$: the data points are calculated with a flat pedestal; the line is based on $v_2$ subtracted yields.}
\efig

\bfig
  \includegraphics[width=0.75\linewidth,trim=0 8 0 5,clip=true]{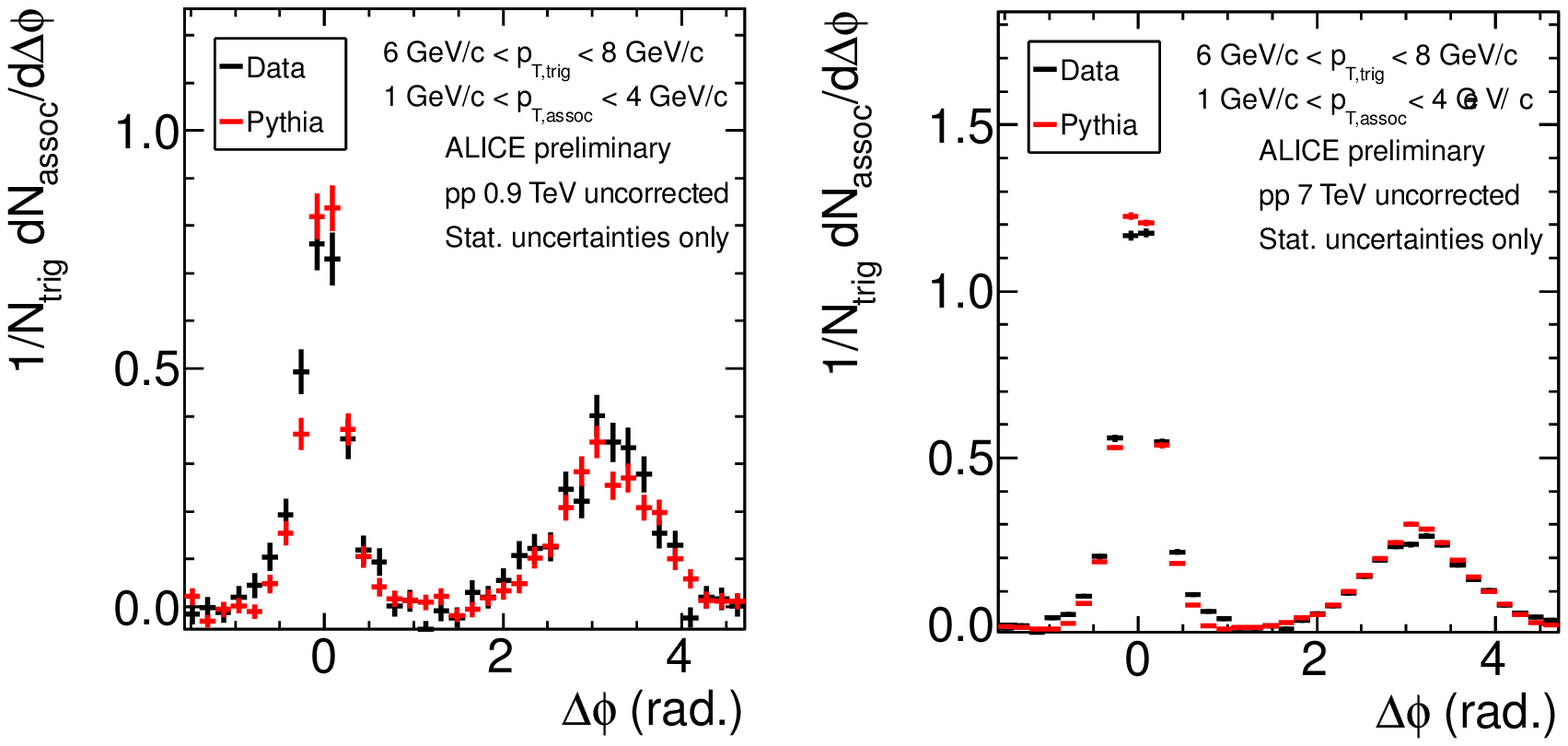}
  \caption{\label{fig_ppref} Uncorrected pedestal-subtracted per-trigger yields from pp collisions at 0.9 (left) and \unit[7]{TeV} (right) are compared to a scaled MC (Pythia 6.4 with the tune Perugia-0).}
\efig

To quantify the effect of the in-medium energy loss, ratios of central to peripheral yields are calculated $\icp = Y_{\rm central} / Y_{\rm peripheral}$
where $Y_{\rm central}$ ($Y_{\rm peripheral}$) is the yield in central (peripheral) collisions, respectively. \figref{fig_icp} shows $\icp$ using the flat pedestal (data points) and $v_2$ subtracted yields (lines). 
That the only significant difference is in the lowest bin of $\pta$ confirms the small influence of flow in this $\pt$ region. It should be noted that we only consider $v_2$ here, although the $v_3$ contribution might be of the same order, particularly for central events. The away-side suppression from in-medium energy loss is seen, as expected. Moreover, there is an unexpected enhancement above unity on the near-side.

To study this further, and in particular if the enhancement is due to using peripheral events in the denominator, it is interesting to calculate $\iaa = Y_{\rm Pb-Pb} / Y_{\rm pp}$
where $Y_{\rm Pb-Pb}$ ($Y_{\rm pp}$) is the yield in Pb-Pb (pp) collisions, respectively.
No pp collisions at the same center-of-mass energy than the recorded Pb-Pb collisions had been produced yet at the time of this analysis. Therefore the option of using a MC as reference has been investigated. 
\figref{fig_ppref} compares uncorrected pedestal-subtracted per-trigger yields of pp collisions taken with ALICE to Pythia\cite{pythia6} 6.4 with the tune Perugia-0\cite{perugia0} at $\sqrt{s} =$ 0.9 and \unit[7]{TeV}. The MC has been scaled such that the yields on the near side agree with each other. The required scaling factor is $0.8 - 1$ depending on $\pt$. One can see that the away side is described well without applying an additional scaling. The scaling factor interpolated to the Pb-Pb energy of 2.76 TeV is then found to be $0.93 \pm 13\%$.

Yields extracted from Pythia 6.4 Perugia-0 with the mentioned scaling factor are used to measure $\iaapythia$, shown in \figref{fig_iaapythia}. As before the data points use the flat pedestal subtraction and the lines use the $v_2$ subtraction. The difference is rather small and only in the smallest $\pta$ bins.
$\iaapythia$ in peripheral events is consistent with unity, but the near side is slightly higher than the away side. This could indicate a slightly different description of the near and away side in the MC. The qualitative behavior of $\iaapythia$ in central events is consistent with $\icp$. The away side is suppressed and the near side significantly enhanced. Such an enhancement has not been reported at lower energies. E.g. STAR measured a near-side $\iaa$ consistent with unity\cite{iaastar}.

\bfig
  \includegraphics[width=0.9\linewidth,trim=0 15 0 12,clip=true]{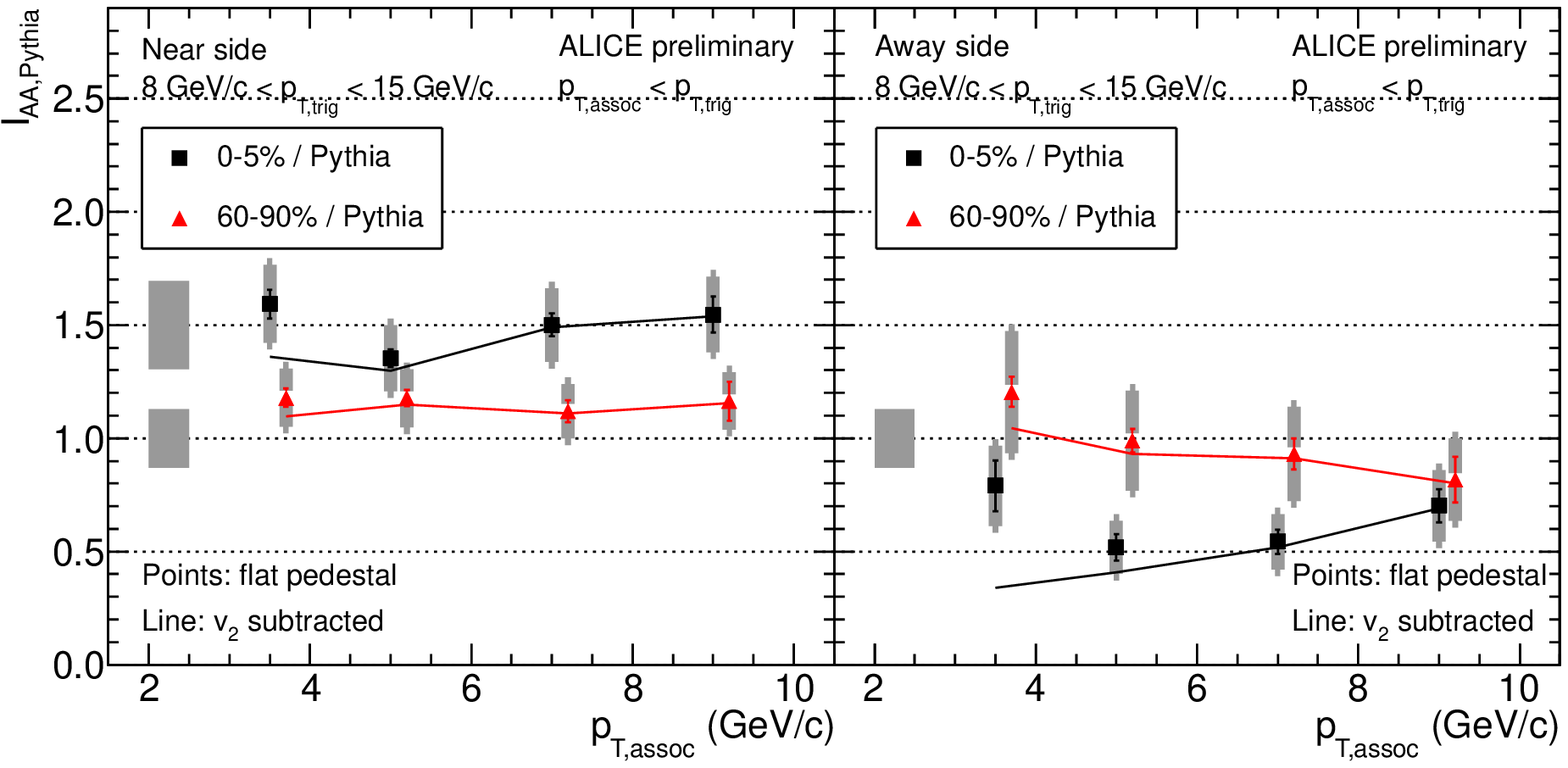}
  \caption{\label{fig_iaapythia} $\iaapythia$: the data points are calculated with a flat pedestal; the line is based on $v_2$ subtracted yields.}
\efig

\vspace{-0.2cm}
\paragraph{Near-Side Enhancement}
A near-side enhancement at LHC was predicted albeit for larger $\ptt$: an enhancement of $10-20\%$ is reported and attributed to the enhanced relative abundance of quarks w.r.t. gluons escaping the medium.\cite{Renk:2007rn} Gluons couple stronger to the medium due to their different color charge and their abundance is reduced. The quarks fragment harder and thus produce an enhanced associated yield. 
Furthermore, a near-side enhancement can be understood if one assumes that the near-side parton is also quenched. Then trigger particles with similar $\pt$ stem from partons with higher $\pt$ in Pb-Pb collisions than in pp collisions. Consequently, more energy is available for particle production on near and away side. 

It should be stressed that a MC was used as a reference for $\iaapythia$ and it will be interesting to study if $\iaa$ using pp collisions shows the same behavior. Such a study is ongoing using newly taken data of pp collisions provided by the LHC in the week after this conference.

\end{document}